# The role of fullerene shell upon stuffed atom polarization potential

M. Ya. Amusia[1,21] and L. V. Chernysheva[2]

[1]*The Racah Institute of Physics, the Hebrew University of Jerusalem, Jerusalem 91904, Israel*
[2]*A. F. Ioffe Physical-Technical Institute, St. Petersburg 194021, Russian Federation*

**Abstract:**
We have demonstrated that the polarization of the fullerene shell considerably alters the polarization potential of an atom, stuffed inside a fullerene. This essentially affects the electron elastic scattering phases as well as corresponding cross-sections. We illustrate the general trend by concrete examples of electron scattering by endohedrals Ne@$C_{60}$ and Ar@$C_{60}$.

To obtain the presented results, we have suggested a simplified approach that permits to incorporate the effect of fullerenes polarizability into the Ne@$C_{60}$ and Ar@$C_{60}$ polarization potential. As a result, we obtained numeric results that show strong variations in shape and magnitudes of scattering phases and cross-sections due to effect of fullerene polarization upon the endohedral polarization potential.

**1**. At first glance, the addition of a single relatively small atom inside a fullerene should not affect the electron elastic scattering cross-section of the latter, since the presence of an additional atom inside alters inessentially the total size of the system under consideration. As it was demonstrated recently in [1] and [2], the quantum interference changes the situation impressively, so that the total partial wave *l* phase $\delta_l^{A@C_N}$ of an electron scattered upon endohedral A@$C_N$ is with good accuracy equal to the sum of scattering phases $\delta_l^A$ and $\delta_l^{C_N}$ of electrons upon atom A, stuffed inside the fullerene $C_N$, and the $C_N$ itself. It means, counterintuitively, that a single atom contribution is quite big as compared to the background of $C_N$ cross-section.

In [1] we have performed calculations, assuming that the incoming electron feels the Hartree-Fock $\hat{V}_{HF}(r)$ potential of the atom A as well as the static $W_F(r)$ and polarization $V_F^{pol}(r)$ potentials of the $C_N$. The inclusion of $V_F^{pol}(r)$ proved to be very important, since $C_N$ is a highly polarizable object, as compared to the atom A.

However, we know that the polarization potential of the atom A modifies essentially its scattering phases and respective cross-sections. Since the atom's A phase is clearly reflected in $\delta_l^{A@C_N}$, one has to investigate the effect of atom A polarization potential $\hat{V}_A^{pol}(r)$ upon $\delta_l^{A@C_N}$. This is why we investigate this effect here. This is the first aim of this Letter.

Moreover, since the fullerene is a highly polarizable object, it can affect the atom's A polarization potential leading instead to a potential that accounts modification of $\hat{V}_A^{pol}(r)$ by the fullerene's shell that we denote as $\hat{V}_{FA}^{pol}(r)$. To investigate the changes that happen when $\hat{V}_{A}^{pol}(r)$ is substituted by $\hat{V}_{FA}^{pol}(r)$ is the second aim of this Letter.

---

[1]amusia@vms.huji.ac.il

As concrete objects of calculations we choose almost ideally spherical fullerene $C_{60}$ and endohedrals Ne@$C_{60}$ and Ar@$C_{60}$ with centrally located quite small and spherical atoms Ne and Ar.

It is in place to remind one general property of the behavior of scattering phases upon a static potential $U$. Let the phases $\delta_l(E)$ as functions of energy $E$ be normalized in such a way that $\delta_l(E \to \infty) \to 0$. If the target consists of electrons and nuclei and exchange between incoming and target electrons is taken into account, an expression $\delta_l(0) = (n_l + q_l)\pi$ proved to be valid. Here $q_l$ is the number of bound electron states with the angular momentum $l$ in the target itself, while the number of bound electron states with angular momentum $l$ in the system $e + U$ is $n_l$ [3, 4].

Therefore, the behavior of phases as functions of $E$ is qualitatively different in cases when we treat the target with and without taking into account the exchange, e.g. in Hartree or Hartree-Fock (HF) approximations. Note that in these two cases the phases deviate from each other (in numbers of $\pi$) although the strength of the potential is almost the same.

In calculations presented below the Ne and Ar atom are treated and polarization potential $\hat{V}_{A}^{pol}(r)$ calculated in the random phase approximation with exchange (RPAE) frame [4], while $C_{60}$ is represented by a static square well potential $W_F(r)$, which parameters are chosen to represent the experimentally known electron affinity of $C_{60}^-$, and low- and medium energy photoionization cross-sections of $C_{60}$ [5]. Along with $W_F(r)$ we take into account the polarization potential $V_F^{pol}(r)$ of the fullerene.

We pay special attention to the development of an approximation that permits to calculate the polarization potential $\hat{V}_{FA}^{pol}(r)$ and corresponding phase-shifts as well as cross-sections.

**2.** In order to obtain electron scattering phases for a spherical endohedral, one has to solve numerically the following equations for radial parts of the one-electron wave functions $P_{El}^{A@C_N}(r)$ [^2]

$$\left( \frac{1}{2}\frac{d^2}{dr^2} + \frac{Z}{r} - \hat{V}_{HF}(r) + W_F(r) - V_F^{pol}(r) - \frac{l(l+1)}{2r^2} - E \right) P_{El}^{A@C_N}(r) = 0. \tag{1}$$

Here Z is the inner atom nuclear charge and $\hat{V}_{HF}(r)$ is the operator of HF non-local potential of the atom A (see definition in e.g. [6]). The asymptotic of $P_{El}^{A}(r)$ determines the scattering phase

---

[^2] We employ the atomic system of units, with electron mass $m$, electron charge $e$, and Planck constant $\hbar$ equal to 1.

$$P_{El}^{A@C_N}(r)|_{r\to\infty} \approx \frac{1}{\sqrt{\pi p}} \sin\left[pr - \frac{\pi l}{2} + \delta_l^{A@C_N}(E)\right]. \qquad (2)$$

Here $p^2 = 2E$.

If one neglects $\left[Z/r - \hat{V}_{HF}(r)\right]$ in (1), the equations (1) and (2) determine scattering function and phase shift of an electron on an empty fullerene, that is denoted as $\delta_l^{C_N}(E)$ and $P_{El}^{C_N}(r)$, respectively, and as "Hartree" on Figures that depict results.

To take into account the atomic polarization potential $\hat{V}_A^{pol}(r)$ or $\hat{V}_{FA}^{pol}(r)$, one has to add one of these potentials to $V_F^{pol}(r)$ in (1), thus obtaining equations for the wave functions $P_{El}^{A@C_{NA}}(r)$ and $P_{El}^{A@C_{NFA}}(r)$ as well as scattering phases $\delta_l^{A@C_{NA}}$ and $\delta_l^{A@C_{N(FA)}}$, respectively.

**3.** More details on how to obtain scattering phases numerically one can find in [6]. The choice of $W_F(r)$ and $V_F^{pol}(r)$ is the same as in [1]: square well for $W_F(r)$ and for $V_F^{pol}(r)$ the following expression $V_F^{pol}(r) = -\alpha_F/2(r^2 + b^2)^2$, where $\alpha_F$ is the static dipole polarizability of a fullerene that for $C_{60}$ and a number of other fullerenes is measured and/or calculated; b is a parameter of the order of the fullerenes radius $R$. This simple version of $V_F^{pol}(r)$ is widely used in atomic scattering calculations (see [7] and references therein).

In principal, the polarization potentials are energy-dependent and non-local. We have an experience to determine it for atoms employing perturbation theory in inter-electron interaction and limiting ourselves by second order perturbation theory in incoming and target electrons interaction (see [4] and references therein).

The equation (1) is then convenient to solve in the integral form and in energy representation, where for partial wave $l$ it looks like (see Chap 3 of [4] and references therein):

$$\left\langle E\ell | \hat{\bar{\Sigma}}^l(E_1) | E'\ell \right\rangle = \left\langle E\ell | \hat{\Sigma}^l(E_1) | E'\ell \right\rangle + \sum_{E''} \left\langle E\ell | \hat{\Sigma}^l(E_1) | E''\ell \right\rangle \frac{1}{E_1 - E'' + i\delta} \left\langle E''\ell | \hat{\bar{\Sigma}}^l(E_1) | E'\ell \right\rangle, \qquad (3)$$

where the sum over $E''$ includes also integration over continuous spectrum.

The polarization interaction $\hat{\Sigma}(E)$ leads to an additional scattering phase shift $\Delta\delta_l(E)$ that is connected to the diagonal matrix element of (3):

$$e^{i\Delta\delta_l(E)} \sin \Delta\delta_l(E) = <E\ell \|\hat{\bar{\Sigma}}^l(E)\| E\ell>. \qquad (4)$$

Instead of semi-empirical potentials, we employ the many-body theory approach with its diagrammatic technique [8, 4]. The matrix elements $\left\langle E\ell | \hat{\Sigma}^l(E_1) | E'\ell \right\rangle$ have the name "irreducible self-energy part of the one-electron Green's function" [8]. This approach accounts for non-locality and energy dependence of the polarization interaction, but to be accurate enough require inclusion of sufficient number of diagrams' sequences.

It appeared, however, that at low incoming electron energies, where the polarization interaction is particularly important, in constructing $\langle E\ell | \hat{\Sigma}'(E_1) | E'\ell \rangle$ it is sufficient to take into account four diagrams [4] presented by (5). These diagrams automatically include some infinite series in electron-vacancy interaction (See Chap. 3 in [4]). A line, directed to the right (left), denotes electron (vacancy).

$$\begin{array}{c}\text{[Diagrams]}\end{array} \tag{5}$$

The wavy line stands for the interelectron interaction. In (5), we use the following notations: $\nu_i = E_i l_i$.

**4.** When we consider an electron colliding with an endohedral, one has to take into account the contribution of the interaction between atomic and fullerenes electrons. Diagrams (6) present examples of such interaction:

$$\begin{array}{c}\text{[Diagrams]} + \text{time reverse terms}\end{array} \tag{6}$$

Here $F$ denotes the fullerenes shell virtual excitations.

Considering the insertion of fullerenes shell virtual excitation and estimating the corresponding contributions, one has to have in mind that between the essential for the scattering process projectile distance $r_p$, the fullerene radius, $R_F$ and the atomic radius, $r_A$ the following inequality exists $r_p > R_C > r_A$. To simplify the problem of taking into account the mutual influence of atomic and fullerenes electron, we enforce this inequality into $r_p \gg R_C \gg r_A$. This permits to limit ourselves by correcting the dipole interelectron interaction only, substituting the Coulomb interelectron potential in the following way $1/|\mathbf{r}_1 - \mathbf{r}_2| \to \mathbf{r}_1 \mathbf{r}_2 / r_2^3$ for $r_1 \ll r_2$.

The variation of the long-range dipole interelectron interaction $V_1$ matrix elements is taken into account similarly to the inclusion of the polarization factor in photoionization of endohedrals as it was demonstrated in [9]. So, we correct them approximately, substituting $|V_1|^2$ by

$$|V_1|^2 \to \left|V_1\left[1-\alpha_F(E_{v_1}-E_{v_4})/R_F^3\right]\right|^2 \qquad (7)$$

in the first and third matrix elements of (5). The alteration of the exchange second and fourth terms of (5) requires substitution:

$$|V_1|^2 \to \left|V_1\left[1-\alpha_F(E_{v_1}-E_{v_4})/R_F^3\right]V_1\left[1-\alpha_F(E_{v_1}-E_{v_2})/R_F^3\right]\right| \qquad (8)$$

We left unchanged other than dipole components of interaction matrix elements.

**5**. To perform calculations, we have to choose concrete values for the $C_{60}$ potentials. The potential $W_F(r)$ is represented by a potential well with the depth 0.52 and inner $R_1$ (outer $R_2$) radiuses equal to $R_1 = 5.26$ ($R_2 = 8.17$). Note that $R_F = (R_1+R_2)/2$. In [9] (see also in [4]) we have calculated the polarizability $\alpha_F(\omega)$. Details on how to find $\langle E\ell|\hat{\Sigma}^I(E_1)|E'\ell\rangle$ and to solve equations (3) and (4) one can find in Chap. 3 of [4].

In Fig. 1 we illustrate the results of calculations of the scattering phases and cross-sections by the cases of Ne@$C_{60}$ and Ar@$C_{60}$ 3. In Fig. 1 we present data for the *s*- phases $\delta$ and their contribution to the cross-sections $\sigma$ of electron scattering upon Ar and Ar@$C_{60}$. The curves $\delta^{Ar}$ and $\sigma^{Ar}$ represent data for $e+Ar$ collision that took into account the action of polarization interaction (5). Data $\delta^F$ and $\sigma^F$ for $e+C_{60}$ are results of solving (1) neglecting the term $\left[Z/r - \hat{V}_{HF}(r)\right]$. Results of RPAE calculations $\delta^{Ar@C_{60}}$ and $\sigma^{Ar@C_{60}}$ for $e+Ar@C_{60}$ mean combination of calculating separately polarization interaction for isolated atom Ar and solving (1). RPAE$_F$ denotes results for $\delta^{Ar@C_{60F}}$ and $\sigma^{Ar@C_{60F}}$ obtained after solving (3), with account of (5). Here all intermediate states $v_2, v_3, v_4$ are solutions of (1). The curves denoted as RPAE$_{FA}$ are results for $\delta^{Ar@C_{60FA}}$ and $\sigma^{Ar@C_{60FA}}$ of calculations similar to the case of RPAE$_F$, but with polarization interaction, that along with (5) includes by using (7) and (8), also diagrams exemplified by (6). Fig. 2 presents similar results, but for *p*-wave of the electron scattering with Ne and Ne@$C_{60}$.

It is remarkable that the cross-section of $e+C_{60}$ collision rapidly drops down with electron energy growth for both presented in Fig. 1 and 2 phases. Starting from 0.2 Ry the contribution of stuffed atom A became comparable or even bigger than that of the $C_{60}$ itself.

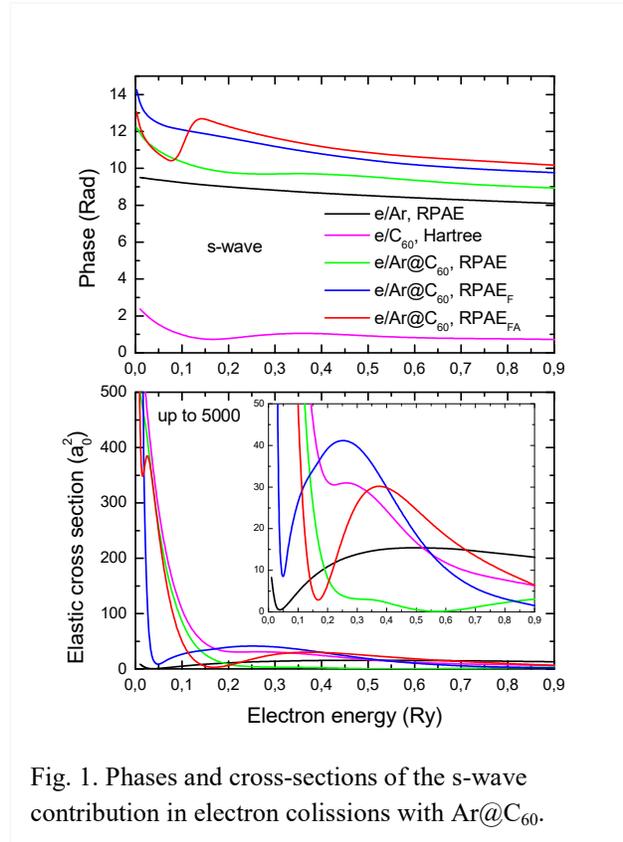

Fig. 1. Phases and cross-sections of the s-wave contribution in electron colissions with Ar@$C_{60}$.

We see that the property of additivity discussed in [1] and [2] is accurate if the effect of the fullerene shell upon atomic polarization interaction presented by (5) is small. It appeared, however that already inclusion of polarization interaction (5), but with endohedral wave functions, considerably affects the scattering phases and cross-sections.

Fig. 3 presents the total elastic electron scattering cross-sections upon Ne@$C_{60}$ and Ar@$C_{60}$. We have calculated first three scattering phases – *s*-, *p*- and *d*. For considered energies, particularly below 0.3 Ry it is enough.

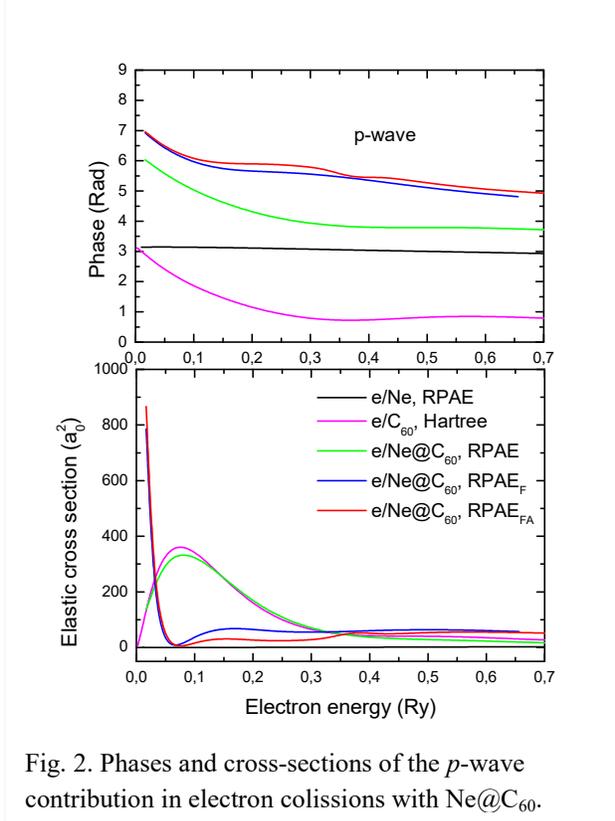

Fig. 2. Phases and cross-sections of the *p*-wave contribution in electron colisions with Ne@$C_{60}$.

Note that as is evident from Fig. 3, each step in increasing the accuracy of our approach leads to prominent changes in the cross-section.

We see that due to presence of the inner atom, the cross-sections acquire a very big resonance at low energy, perhaps even at $E \to 0$. At $E \approx 0.1$ Ry the cross-sections have a deep minimum that in atomic scattering is called Ramsauer minimum. This minimum appears only in $RPAE_F$ and $RPAE_{FA}$, but does not exist in RPAE.

**6.** Using concrete examples we have demonstrated that the elastic scattering of electrons upon endohedrals is an entirely quantum mechanical process, where addition of even a single atom can qualitatively alter the multi-particle cross-section.

Even the crudest account of the fullerene influence upon the caged atom polarization potential (that is achieved by using endohedral's electron wave functions (1) to describe the intermediate states $v_2, v_3, v_4$ in (5)), alters the phases and cross-sections impressively (compare results for RPAE and $RPAE_F$ in Fig 1-3).

Surprisingly enough, the fullerenes shell dynamic polarization strongly modifies the stuffed atom polarization potential. This is clarified by taking into account the corrections (5) (compare results for RPAE and $RPAE_F$).

It is essential that the inner atom's polarization potential is strongly modified

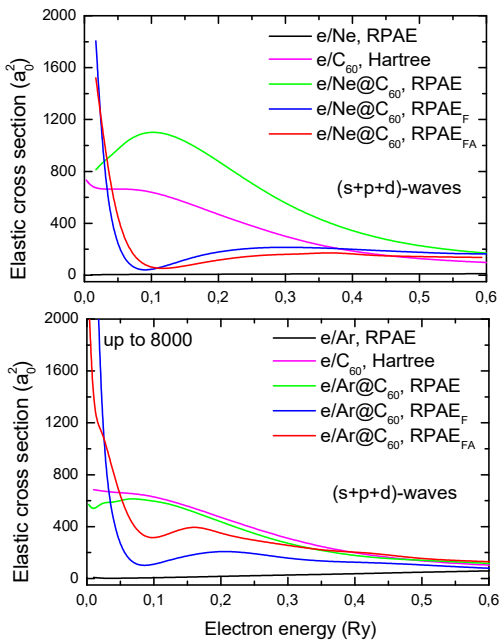

Fig. 3. Total cross-section of electron elastic scattering upon Ne@$C_{60}$ and Ar@$C_{60}$.

under the influence of the fullerene's shell polarizability that is described using (6-8).

Only further clarification of the polarization potential can permit to make a decisive conclusion: either a low-energy scattering resonance or of an extra bound state between an incoming electron and an endohedral exist.

We do believe that the presented results will stimulate theoretical and experimental research of low-energy elastic scattering of electrons by endohedral atoms.

**Figures** (color on line)

**Figure captions**

Fig. 1. Phases and cross-sections of the s-wave contribution in electron colissions with Ar@$C_{60}$.
Fig. 2. Phases and cross-sections of the *p*-wave contribution in electron colissions with Ne@$C_{60}$.
Fig. 3. Total cross-section of electron elastic scattering upon Ne@$C_{60}$ and Ar@$C_{60}$.

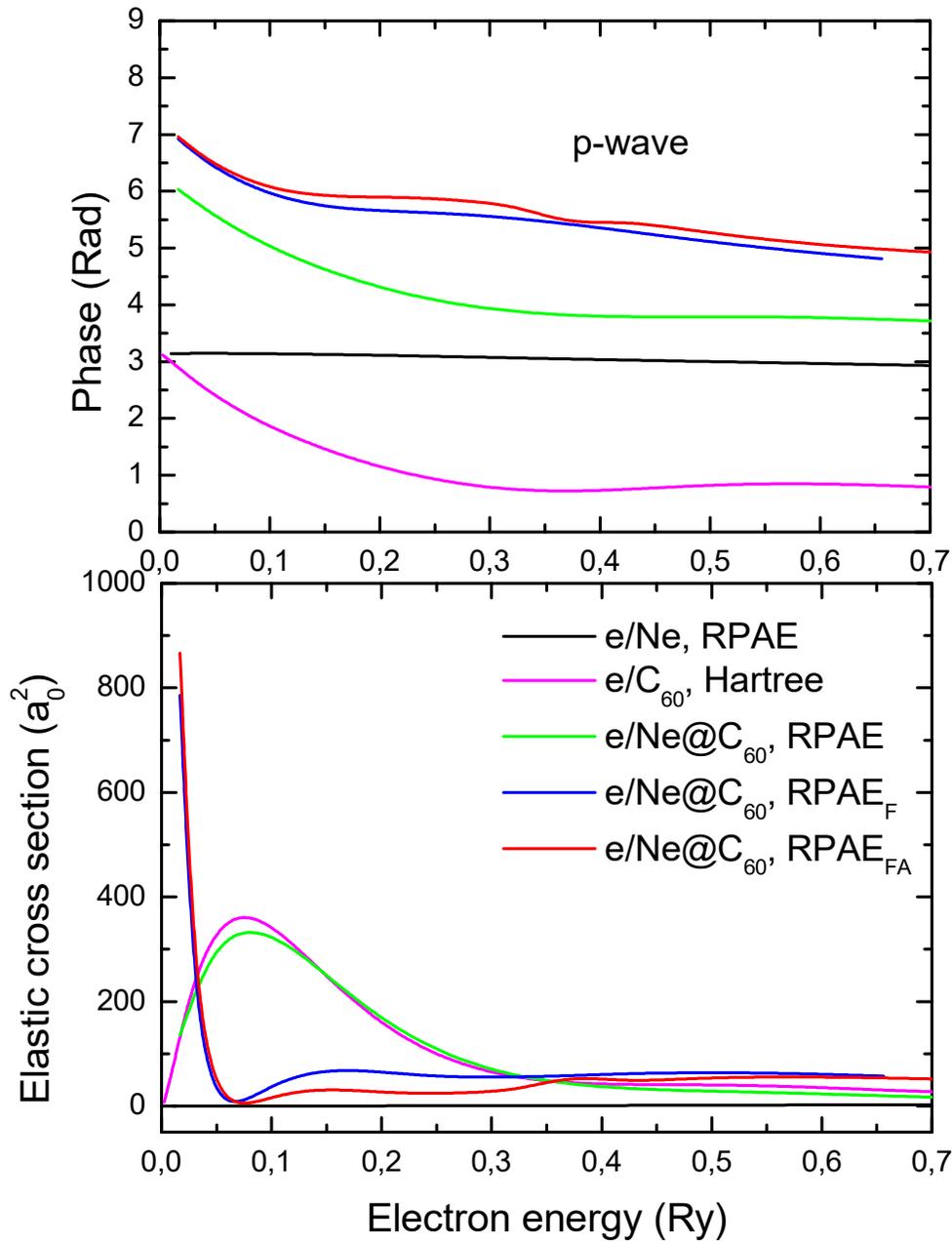

Fig. 1. Phases and cross-sections of the s-wave contribution in electron colissions with Ar@C$_{60}$.

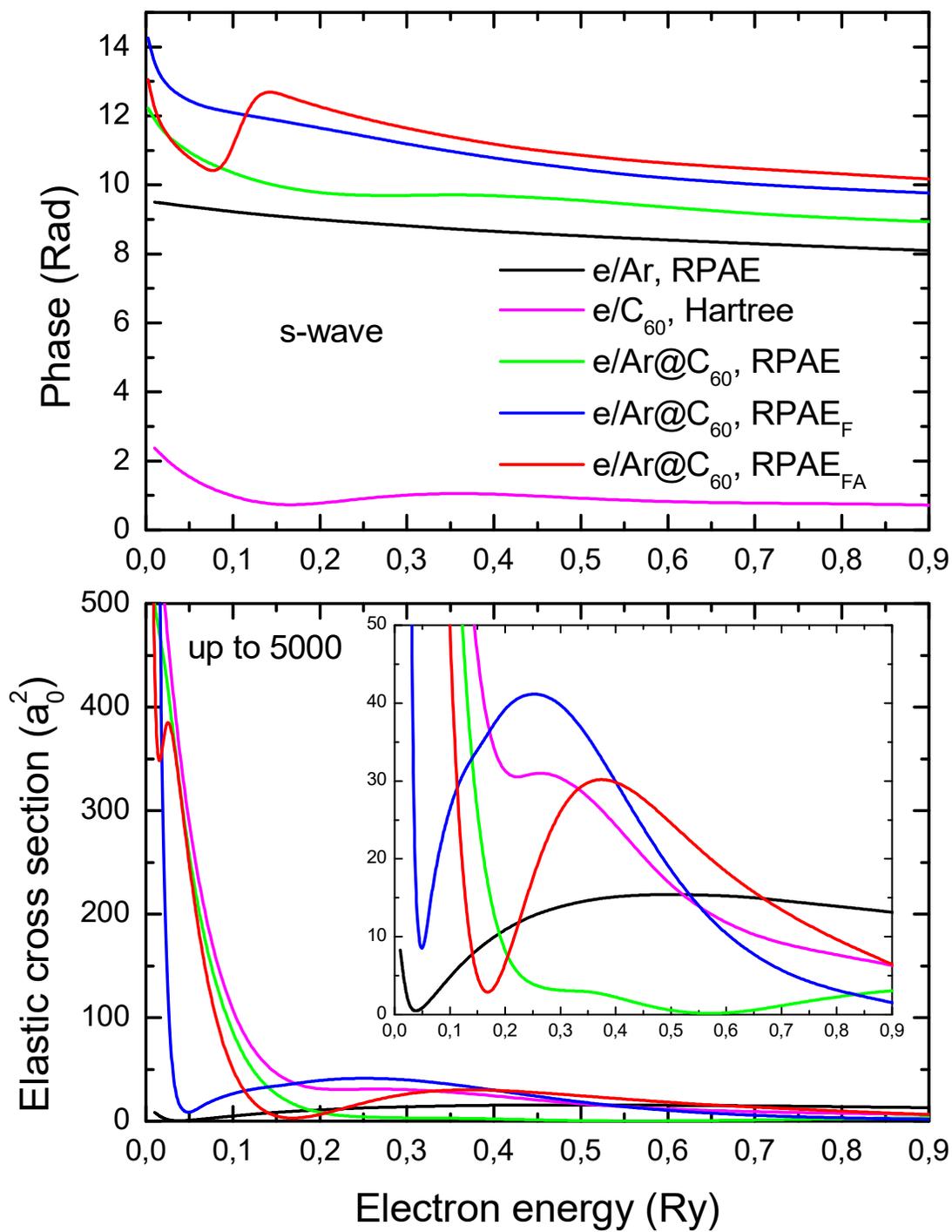

Fig. 2. Phases and cross-sections of the *p*-wave contribution in electron colisions with Ne@C$_{60}$.

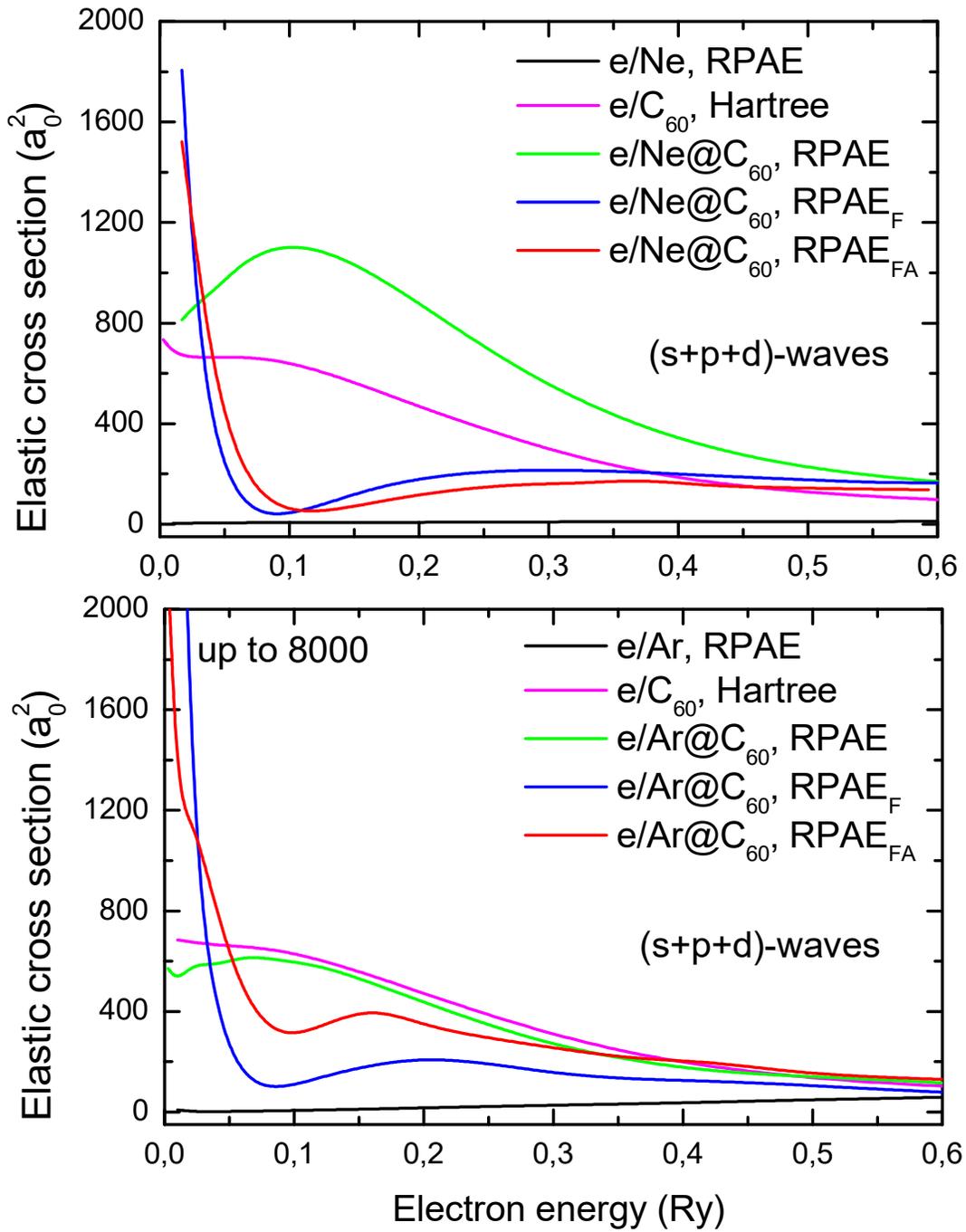

Fig. 3. Total cross-section of electron elastic scattering upon Ne@$C_{60}$ and Ar@$C_{60}$.